\documentclass[conference]{IEEEtran}
\ifCLASSINFOpdf
   \usepackage[pdftex]{graphicx}
\else
\fi
\usepackage{url}
\usepackage{lipsum}
\usepackage{placeins} 	

\begin{document}
%
\title{Hidden-Markov-Model \\ Based Speech Enhancement}

\author{\IEEEauthorblockN{Daniel Dzibela,
Armin Sehr}
\IEEEauthorblockA{Faculty of Electrical Engineering and Information Technology\\
Ostbayerische Technische Hochschule Regensburg\\
93049 Regensburg, Germany\\}
\texttt{daniel.dzibela@st.oth-regensburg.de},\,\,\, \texttt{armin.sehr@oth-regensburg.de}
}


%


\maketitle

\begin{abstract}
The goal of this contribution is to use a parametric speech synthesis system for reducing background noise and other interferences from recorded speech signals. In a first step, Hidden Markov Models of the synthesis system are trained. \\
Two adequate training corpora consisting of text and corresponding speech files have been set up and cleared of various faults, including inaudible utterances or incorrect assignments between audio and text data. Those are tested and compared against each other regarding e.g. flaws in the synthesized speech, it's naturalness and intelligibility. Thus different voices have been synthesized, whose quality depends less on the number of training samples used, but much more on the cleanliness and signal-to-noise ratio of those. Generalized voice models have been used for synthesis and the results greatly differ between the two speech corpora. \\
Tests regarding the adaptation to different speakers show that a resemblance to the original speaker is audible throughout all recordings, yet the synthesized voices sound robotic and unnatural in smaller parts. The spoken text, however, is usually intelligible, which shows that the models are working well. \\
In a novel approach, speech is synthesized using side information of the original audio signal, particularly the pitch frequency. Results show an increase of speech quality and intelligibility in comparison to speech synthesized solely from text, up to the point of being nearly indistinguishable from the original. \\
\end{abstract}


%
\IEEEpeerreviewmaketitle

\section{Introduction}

Acoustic human-machine-communication, e.g. speech recognition and speech synthesis, is becoming increasingly popular nowadays due to the growing number of people owning smartphones. With these it is possible to ask questions and get them answered directly by speech on both sides. Achieving word error rates of less than 5\,\% is, apart from other factors, owed to speech enhancement as well \cite{accuracy}. \\
Speech enhancement aims at improving the quality and/or the intelligibility of speech signals by reducing unwanted noise and other interferences, like competing speakers or reverberation \cite{Benesty}. \\
In a novel approach however, speech synthesis is utilized for this task as well. With a combination of both methods better results can be achieved under certain conditions.
Additionally, side information of the original signal is used to improve the synthesized signal's quality and thus increase the intelligibility even more.

\section{The Process of Speech Synthesis}
\label{sec:synthesis}

\begin{figure}[!t]
\centering
\includegraphics[width=3.5in]{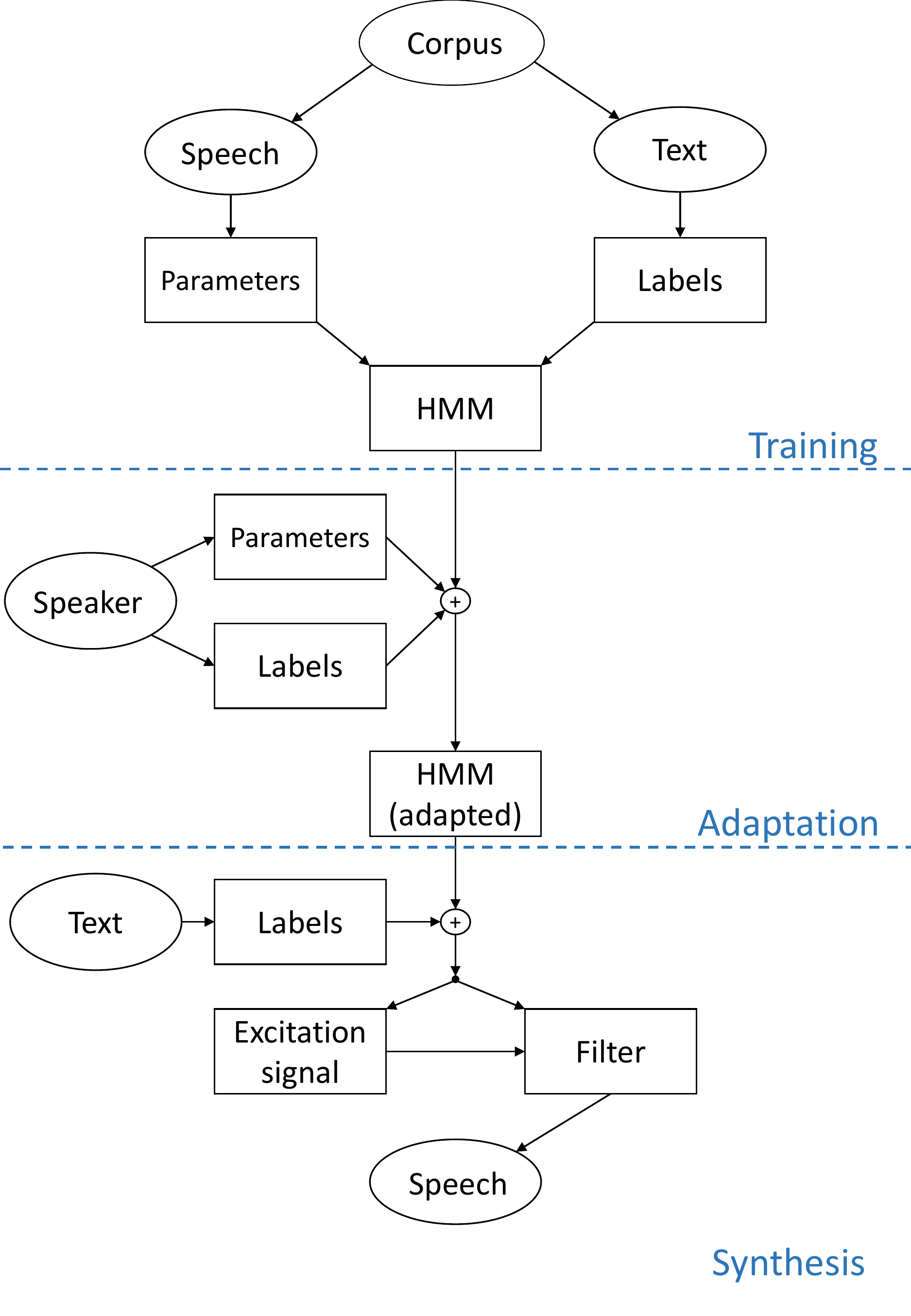}
\caption{Process of parametric speech synthesis}
\label{synthese}
\end{figure}

First the process of speech synthesis has to be described. A short overview is given in Figure \ref{synthese}. \\
The data required for the training is a large speech corpus consisting of usually tens of hours of audio data from many different speakers \cite{zen_specom}. Additionally, the corresponding text files are given, so that it is exactly known what is spoken. From these, labels describing the text and parameters characterizing the language can then be extracted with which Hidden-Markov-Models (HMMs) are trained. For each phoneme, i.e. each sound in a language, several HMMs are created, which link speech to the parameters of the source-filter-model used for synthesis.\\
Generally speaking the human voice can be described as an excitation signal, which is convoluted with a filter \cite{slides}. The signal, which models the vocal chords, is either noise for unvoiced phonemes, mainly consonants, or a pulse train for voiced phonemes like many vowels. The pulse train has a specific frequency $F_0$, which varies in different speakers. This excitation signal is then convoluted with a filter, which represents the oral and nasal cavity. For each phoneme the position of tongue, jaw etc. are different, resulting in a different filter for each one. \\
However, not only a single phoneme is analyzed, but rather a sequence of three to five, as the context strongly influences the actual sound of each one. The corresponding HMMs are therefore called Triphone or Quinphone HMMs. The result of the training process is an average voice model. \\
While synthesis is already possible with this data alone, speaker adaptation is usually performed as well, even though this step is optional. This describes mimicking a specific speaker or his style of speaking. Usually five to seven minutes of data are sufficient for this purpose \cite{tokuda}. In this step, the HMMs' parameters are tuned to better model the adaptation speaker. \\
In the last step, the actual synthesis, only text is required, from which labels are extracted. These are then used in conjunction with the HMMs to select the sequence of parameters for the source-filter-model, which best represent the phonemes spoken.

\section{Evaluation of speech corpora}
In the next section the evaluation of two speech corpora, namely the WSJ0 and the CSTR VCTK, is discussed \cite{WSJ0} \cite{VCTK}. For this purpose, only speech synthesized with the speaker-independent models is compared. The speaker-adapted models are assessed independently, since they used different speakers. It should be noted, however, that there is a second algorithm, designed for improving upon these, called speaker-adaptive training (SAT)\footnote{\,\,The SAT-algorithm also delivers speaker-independent models.} \cite{sat_proposal}. 
Results showed a drastic increase in computation time, about five times longer than the previous speaker-independent models. Thus, including these was not feasible for larger corpora, as calculation of all models for the full WSJ0 corpus took roughly one month. One reason for this is that the computation takes place in a single process on a single CPU. Thus, the computing time could be drastically reduced if e.g. multicore CPUs were supported.

\subsection{WSJ0 Corpus}
\begin{figure}[!t]
\centering
\includegraphics[width=3.5in]{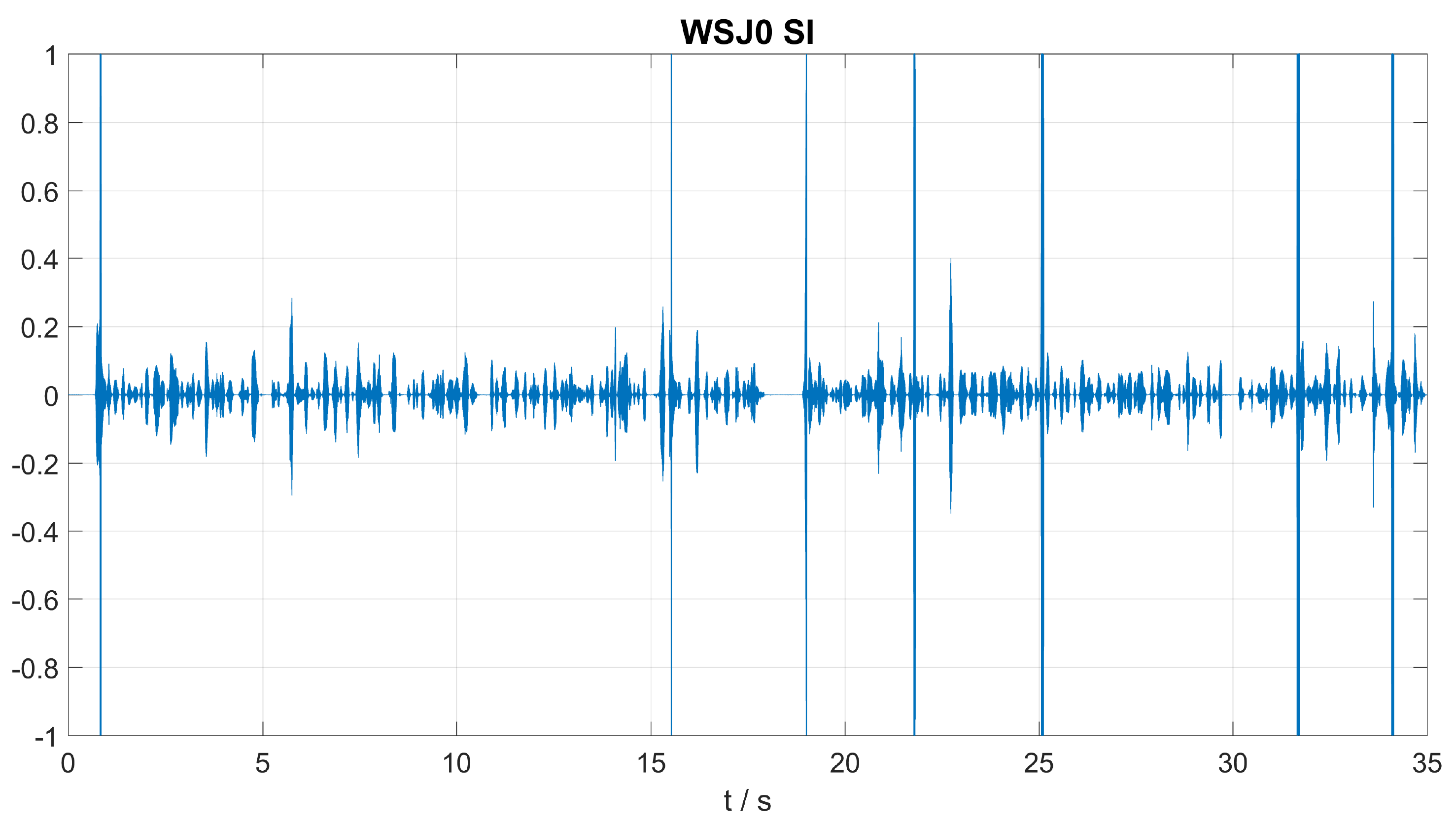}
\caption{Speaker-independent speech synthesized with a training database of approx. 12000 utterances from the WSJ0 corpus}
\label{WSJ0}
\end{figure}

\begin{figure}[!t]
\centering
\includegraphics[width=3.5in]{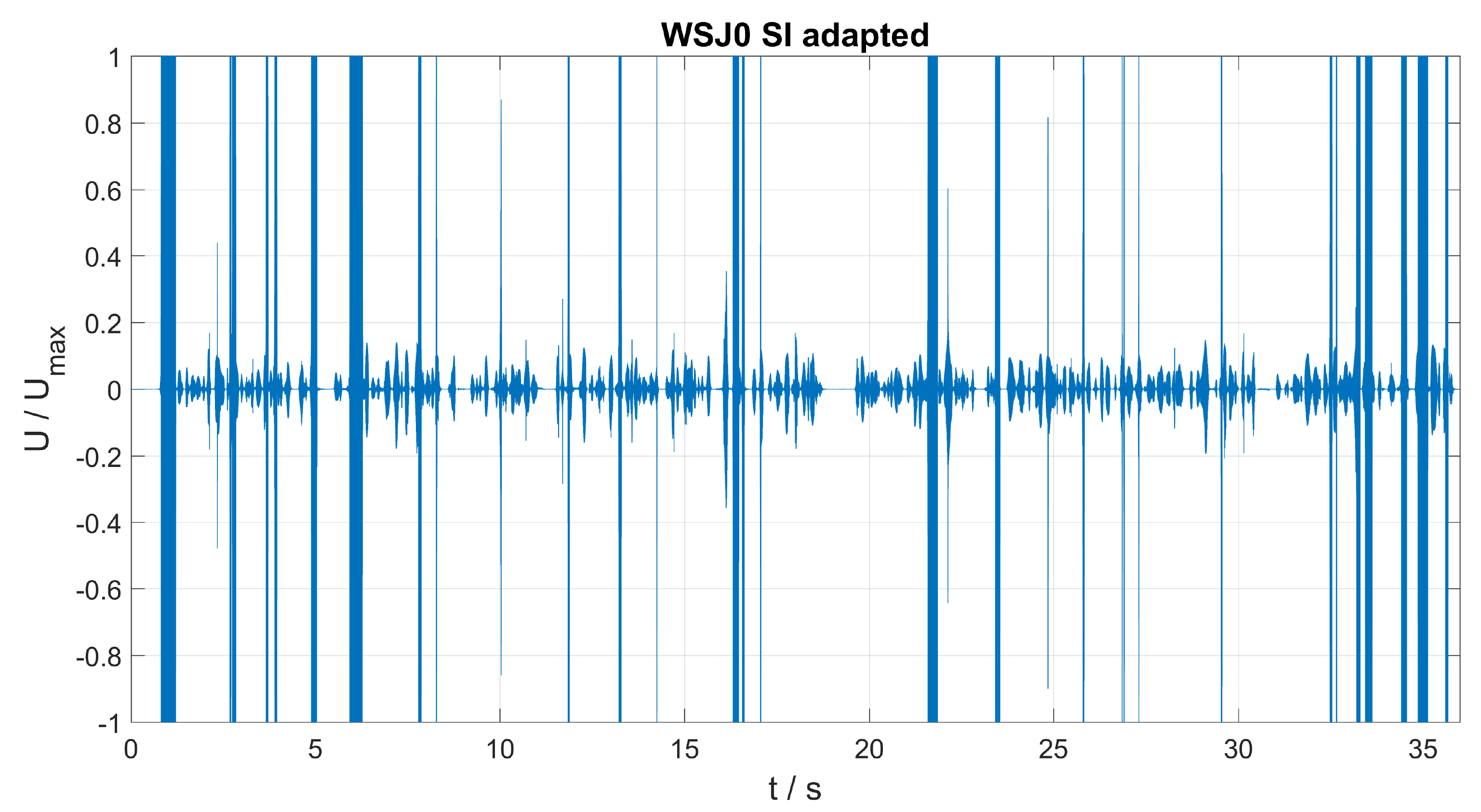}
\caption{Adapted speech synthesized with an independent training database of approx. 12000 utterances from the WSJ0 corpus}
\label{WSJ0_adapted}
\end{figure}

The WSJ0 corpus is the first one to be analyzed. It's name derives from the Wall Street Journal where most of the prompts, i.e. the texts to be read, are drawn from. It should be noted, however, that it was mainly designed for testing speech recognition systems instead of speech synthesis systems \cite{WSJ0_doc}. While it is approximately evenly distributed regarding gender, for 73\,\% of the speakers neither age nor accent is given, making it difficult to choose an adequate training subset. \\
In order to be able to synthesize speech with the corpus, it first had to be cleared of various flaws like mismatch of audio and label files, empty labels or faulty audio data containing accidental sounds. As first tests with a very small subset resulted in poor quality, the full corpus containing approx. 12000 sentences, which is equivalent to approx. 20\,h of audio files, was used. Figure \ref{WSJ0} shows the results. \\
The synthesized speech does sound comprehensible, however the general quality is rather poor. This has several reasons: most of the audio is quite low, while some parts are heavily distorted. Those are mainly found, whenever sibilants like s or sh are uttered, and, as they contain high frequencies, sound very unpleasant. Their occurrence is likely caused by incorrect coefficients for the MLSA (Mel Log Spectrum Approximation) filter. In turn, the audio cannot be amplified either, as the audio would be clipped otherwise. Furthermore the voice sounds unnatural and robotic due to random drops of the pitch frequency. The temporal structure of the sentence is indeed natural. \\
The adapted signals, however, are full of the aforementioned spikes and thus not applicable in any way. Figure \ref{WSJ0_adapted} shows a characteristic example. The sudden increase of faults is probably caused by the low quality and signal-to-noise ratio of the speaker-specific data. In parts with no errors present the adaptation worked well, though. The pitch was adapted as well as the temporal structure.

%
%

\subsection{CSTR VCTK Corpus}
\begin{figure}[!t]
\centering
\includegraphics[width=3.5in]{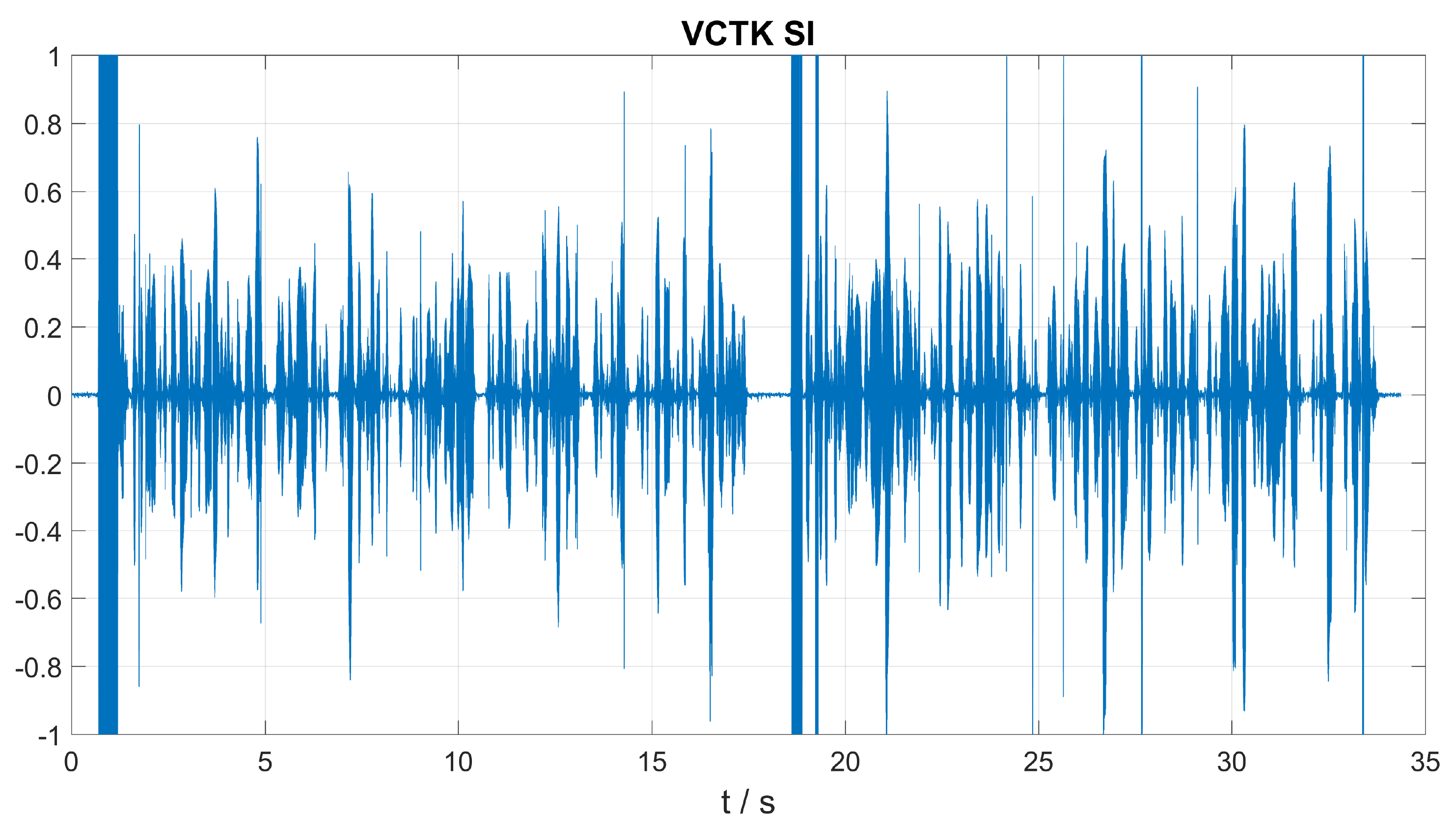}
\caption{Speaker-independent synthesized speech with a training database of approx. 3000 utterances from the CSTR VCTK corpus}
\label{VCTK}
\end{figure}

\begin{figure}[!t]
\centering
\includegraphics[width=3.5in]{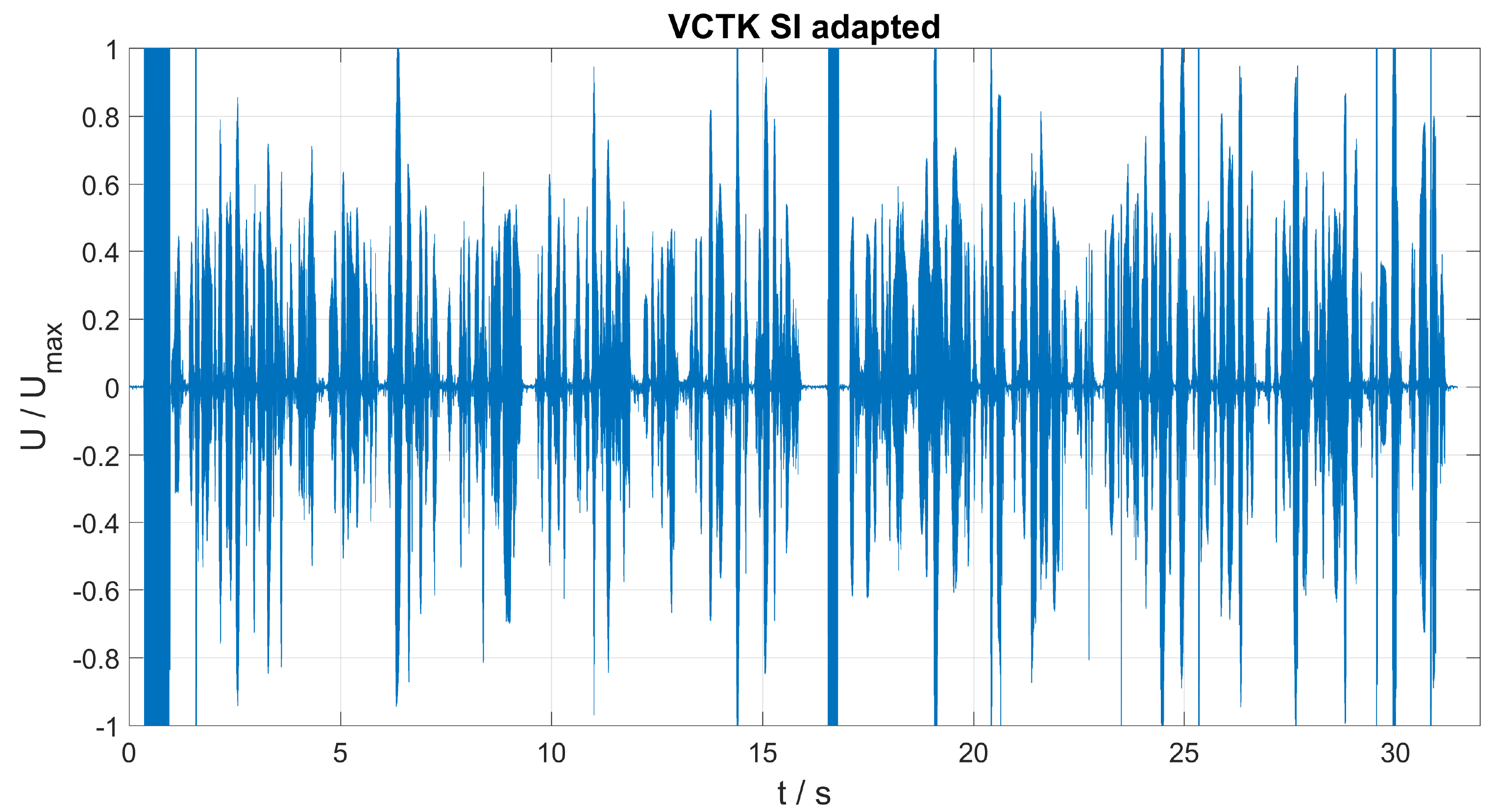}
\caption{Adapted speech synthesized with an independent training database of approx. 3000 utterances from the VCTK corpus}
\label{VCTK_adapted}
\end{figure}

The CSTR VCTK corpus was recorded specifically for HMM-based speech synthesis. It is mainly based on newspaper texts and contains approx. 400 sentences by 109 native English speakers respectively. Their distribution of age and accent are illustrated in Tables \ref{age_vctk} and \ref{accents_vctk}. \\
Tests have been run with a subset of approx. 3000 sentences from nine speakers and the results are illustrated in Figure \ref{VCTK}. Even though more data is usually required to build an adequate average voice model, the quality still exceeds the WSJ0 corpus' results with 12000 files \cite{3000_sentences}. One of the reasons is the generally higher volume as the audio files of the corpus are louder as well.
Again, spikes are present. These, as before, occur probably due to incorrect coefficients for the MLSA filter. In this case, however, they appear only at the beginning of each sentence and are low-frequency. The speech, while slightly less comprehensible than before, sounds a lot more natural, even though the drops in pitch are still present. However, the reason for this might simply be the smaller dataset. Thus, the voice model could not generalize parameters as well and is not as abstract as a model with a larger database. \\
Figure \ref{VCTK_adapted} shows a characteristic, adapted speech signal. While the spikes are still present, their number did not increase. This further proves that the adaptation data from the WSJ0 corpus was faulty. Compared to the speaker-independent results the voice sounds more natural and human, having adapted to the fast speaking rate and the pitch of the speaker. Due to the increased speed, however, the intelligibility slightly decreased for listeners with English as a second language.
\\ All in all the quality of the CSTR VCTK corpus significantly exceeds the one by the WSJ0 corpus.

\begin{table}[!t]
\caption{Distribution of age in the CSTR VCTK corpus}
\label{age_vctk}
\centering
\begin{tabular}{|c|c|}
\hline
    \textbf{Age\,/\,a} & \textbf{Quantity}\\ \hline
    20-29 & 91 \\ \hline
    18-19 & 14 \\ \hline
    30-39 & 3 \\ \hline
    Unknown & 1 \\ \hline
\end{tabular}
\end{table}

\begin{table}[!t]
\caption{Distribution of accents in the CSTR VCTK corpus}
\label{accents_vctk}
\centering
\begin{tabular}{|c|c|}
\hline
    \textbf{Accent} & \textbf{Quantity}\\ \hline
    English & 33 \\ \hline
    American & 22 \\ \hline
    Scottish & 19 \\ \hline
    Irish & 9 \\ \hline
    Canadian & 8 \\ \hline
    Northern Irish & 6 \\ \hline
    South African & 4 \\ \hline
    Indian & 3 \\ \hline
    Australian & 2 \\ \hline
    Welsh & 1 \\ \hline
    New Zealand & 1 \\ \hline
    Unknown & 1 \\ \hline
\end{tabular}
\end{table}


\section{Speech enhancement via speech synthesis with use of side information}

\subsection{Approach}
\begin{figure}[!t]
\centering
\includegraphics[width=3.5in]{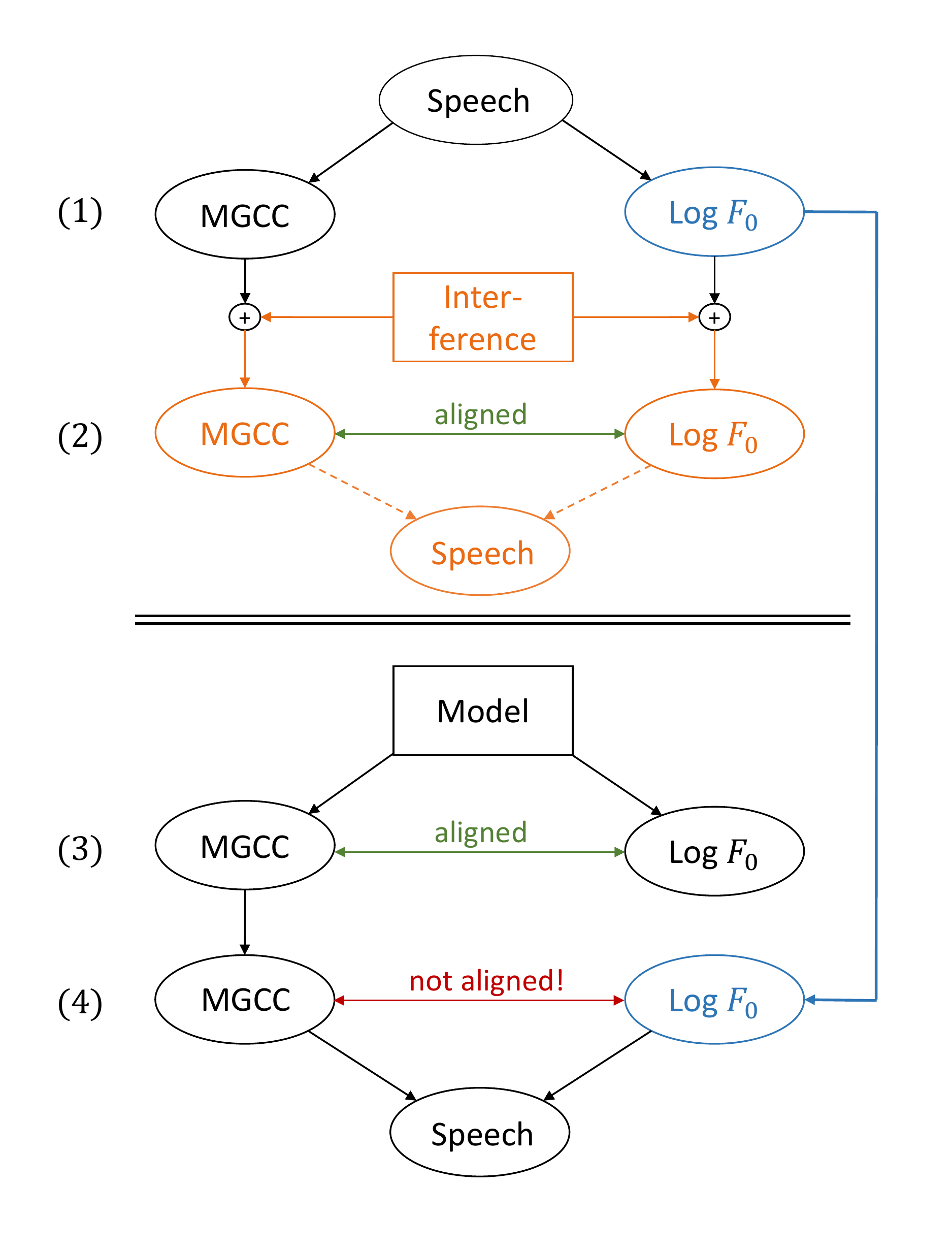}
\caption{Use of side information to enhance synthesized speech}
\label{alignment}
\end{figure}

In a novel approach, HMM-based speech synthesis is used for speech enhancement. This means that the initially spoken sentence is re-synthesized to enhance the audio. To increase the quality of the generated signal, side information of the original speech is used. The goal of this test is to examine the best possible quality, as ideal data is used, which is usually not available in practice. In an oracle test, however, this was assumed to be present, in order to find an upper bound of the maximum achievable quality. \\
Fig. \ref{alignment} outlines the procedure. Log $F_0$ describes the logarithmic pitch frequency, i.e. the base frequency of a speaker, whereas MGCC (mel-generalized cepstrum coefficients) depict the coefficients of the MLSA filter, which model the vocal tract of a speaker, giving him his unique sound \cite{echoes_2002}. Those two parts are needed in order to synthesize speech. \\
Step $(1)$ illustrates the oracle, since neither the undisturbed speech signal nor its' components are available in practice. They are included since the ideal Log $F_0$ is used as side information for the synthesis later in order to examine the best possible speech quality. \\
In practice, however, interferences\footnote{\,\,Those interferences are added to the signal, not the coefficients or Log $F_0$ itself. The diagram simply shows the impact on those.} like noise, reverberation or competing speakers are present, which are modeled in step $(2)$. This is the signal that is usually available. When working with real world data, the Log $F_0$ of this disturbed signal would have to be utilized as side information. \\
Step $(3)$ depicts speech synthesis as described in section \ref{sec:synthesis}, using the components generated by a speaker-independent or adapted voice model in order to create speech. \\
The actual approach is displayed in step $(4)$. The MGCCs of the voice model are used in combination with the Log $F_0$, i.e. the side information, of the original signal. Thus, the original speech signal is re-synthesized and enhanced. As this is an oracle test, ideal side information is used instead of disturbed. \\
The problem, however, is the incorrect alignment of MGCCs from the model and the Log $F_0$ from the speech signal. This means that e.g. the excitation signal for one phoneme, but the coefficients for a different one are used, resulting in faulty speech. To avoid this mismatch, forced alignment of the two components is applied. For this purpose, labels are given to the model which then forcefully tries to find their temporal appearance in the audio and adds them to the label, similar to timestamps. Thus, coefficients can be generated which match the course of the pitch frequency. If interferences are present in the audio, however, the forced alignment is negatively affected as well.

\subsection{Results}
\begin{figure}[!t]
\centering
\includegraphics[width=3.5in]{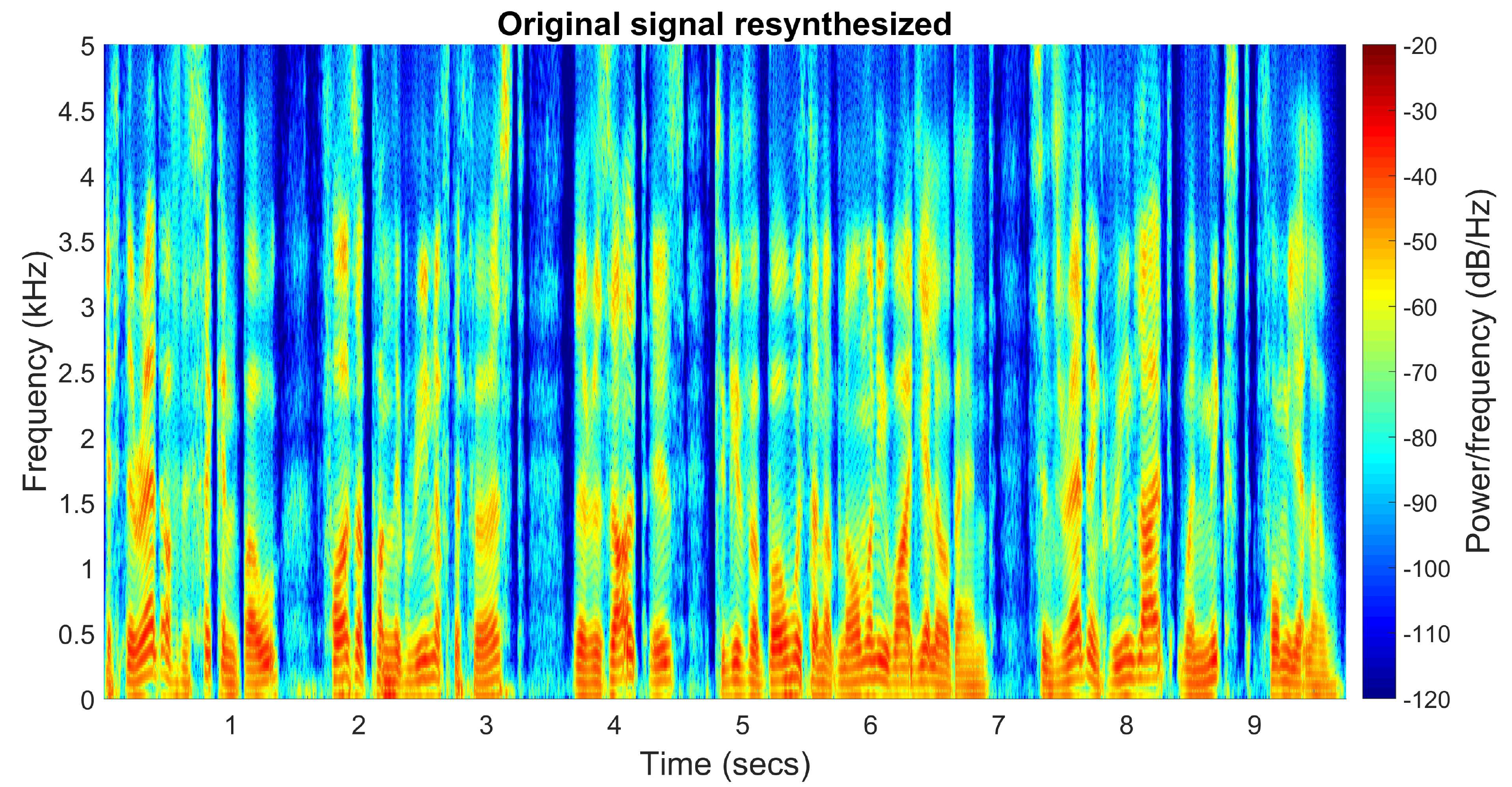}
\caption{Spectrogram of speech resynthesized from original, undisturbed MGCCs and Log $F_0$ as in step $(1)$}
\label{spectrogram_original}
\end{figure}

\begin{figure}[!t]
\centering
\includegraphics[width=3.5in]{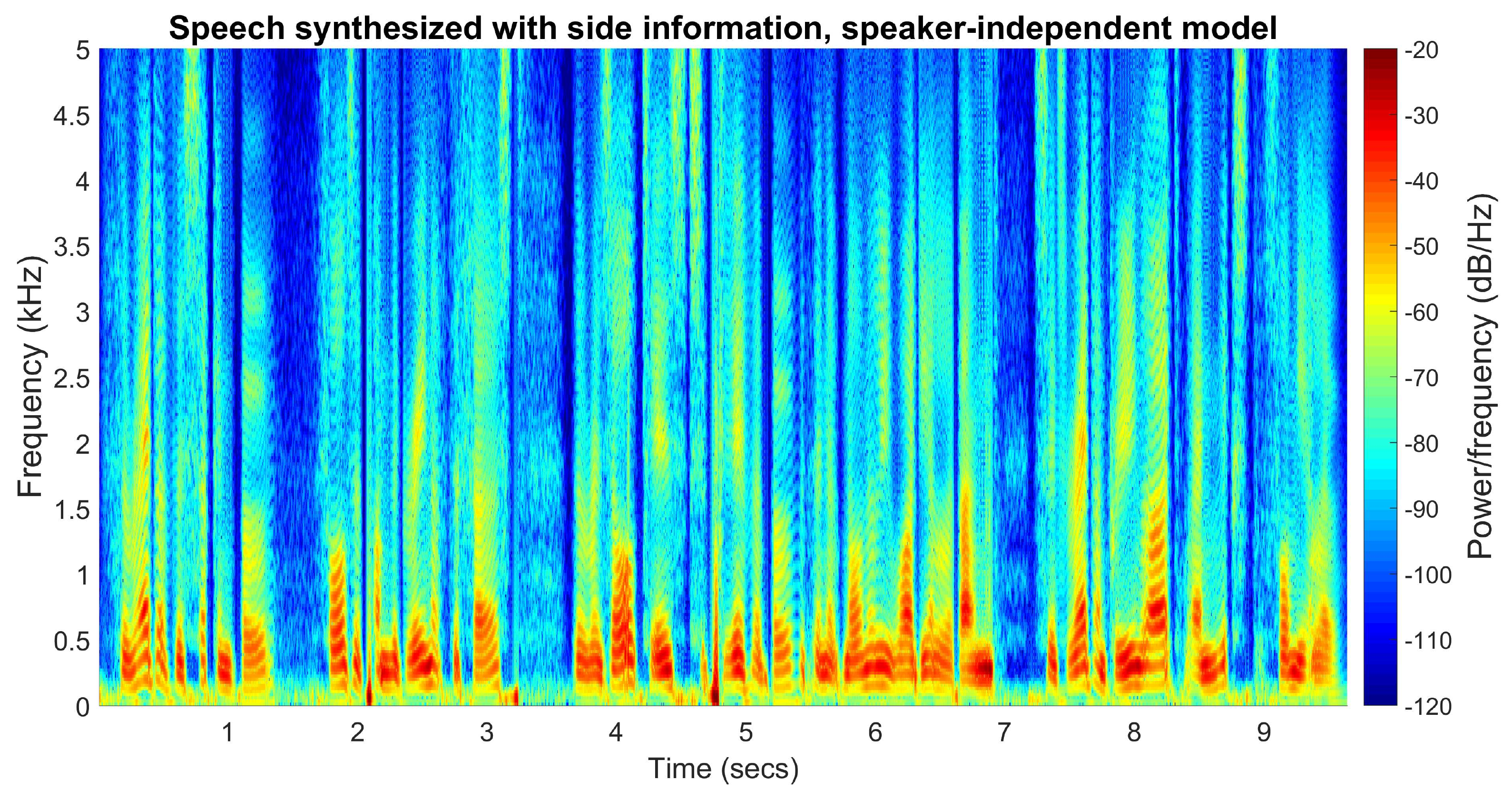}
\caption{Spectrogram of speech synthesized with side information as shown in step $(4)$, using a speaker-independent, average voice model}
\label{spectrogram_side_information}
\end{figure}

\begin{figure}[!t]
\centering
\includegraphics[width=3.5in]{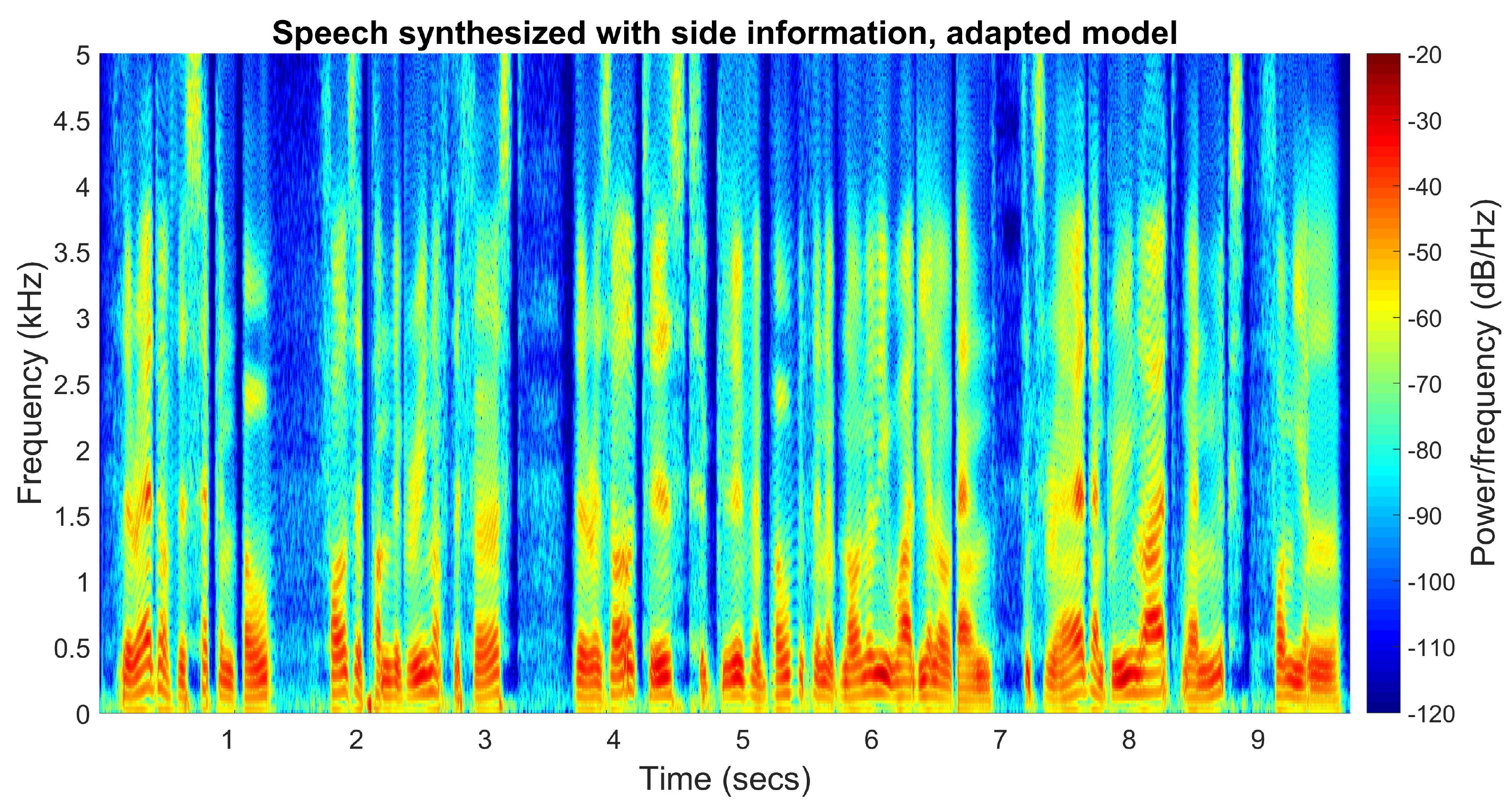}
\caption{Spectrogram of speech synthesized with side information as shown in step $(4)$, using an adapted voice model}
\label{spectrogram_side_information_adapted}
\end{figure}

The signals synthesized from steps $(1)$ and $(4)$ in Fig. \ref{alignment} are now compared in order to examine the best possible quality.
Fig. \ref{spectrogram_original} describes the signal, which is synthesized by the Log $F_0$ and MGCCs from step $(1)$. This illustrates the maximum achievable quality, as there is no voice model involved. \\
Fig. \ref{spectrogram_side_information} pictures the audio generated in step $(4)$ with the speaker-independent, average voice model. In contrast, Fig. \ref{spectrogram_side_information_adapted} represents the signal synthesized using the components of step $(4)$ with the adapted voice model.
All signals have been normalized before computing the spectrogram so that the scale is now equal for all plots. Three major differences are clearly visible: \\
The first is the attenuation of the lowest frequencies in the generated audio data. As their power in the original speech is higher than in the adapted and speaker-independent model, this suggests that high-pass filtering is applied when training the models, as this has shown to improve their accuracy \cite{high-pass-filter}. Since the adapted model comes closer to the spectrogram of the original speaker, it better reflects him in this regard. \\
Furthermore the modulation of the original signal's lower frequencies is much stronger than the side information signals'. This difference is based on the generation process of speech parameters: those are first calculated under dynamic feature constraints, resulting in a very smooth spectrum. Next, advanced amplitude modulation is used to loosen these, which still does not solve the problem completely. The spectrum of natural speech often is rather rough. An over-smoothed spectrum in synthesized speech, however, leads to the voice sounding muffled and dull. So in order to compensate for this, another term is added when calculating the parameters, which penalizes them if they are over-smoothed \cite{slides}. This process does greatly increase the naturalness of the signal, yet still leaves differences to the original signal \cite{hts_2005}. As it is a general problem of speech parameter generation, this difference persists for both voice models.\\
The last and most important difference is the attenuation of higher formants, i.e. the resonant frequencies of the vocal tract, in the ranges of approx. 2.5\,kHz and 3 - 3.5\,kHz \cite{rabiner_schafer}. When comparing Fig. \ref{spectrogram_original} and \ref{spectrogram_side_information} it becomes particularly obvious. This leads to somewhat muffled speech using the speaker-independent model, which does not sound as clear and sharp as the re-synthesized original. This is due to the average model not reflecting the speaker-specific vocal tract, which is represented by the vowel formants \cite{speaker_specific_formants}. The adapted model, however, describes the speaker very well, as can be seen by the similar emphasis of formants in approximately the same frequency ranges as the original. Even though minor differences are still visible in the spectrogram, listening tests showed a very strong resemblance to the original speaker, with a notable increase of speech quality compared to the speaker-independent model. \\
All in all the increase of quality and intelligibility of synthesized speech, which uses side information and the adapted model, is clearly present across the whole signal compared to normally generated speech. In some parts it even comes close to being indistinguishable from the original re-synthesized signal.

\section{Conclusion}

In this paper an evaluation of two speech corpora for HMM-based speech synthesis was presented. As the CSTR VCTK corpus was designed specifically for this task, it proved to be better suited than the WSJ0 corpus, which was created for speech recognition. Furthermore a novel approach to speech enhancement was illustrated, which involves HMM-based speech synthesis and the use of side information. A first test shows very promising results, with a clear increase of quality and intelligibility as compared to normally generated speech. Next, different tests are going to be run examining the impact of each step on the final result.



\bibliographystyle{IEEEtran}

\end{document}